# 14 Market-Oriented Cloud Computing and the Cloudbus Toolkit

Rajkumar Buyya[1,2], Suraj Pandey[1], and Christian Vecchiola[1]

[1] Cloud Computing and Distributed Systems (CLOUDS) Laboratory,
Department of Computing and Information Systems,
The University of Melbourne, Australia

[2] Manjrasoft Pty Ltd, Melbourne, Australia

## TABLE OF CONTENTS

# SUMMARY

**Cloud computing has penetrated the Information Technology industry deep enough to influence major companies to adopt it into their mainstream business. A strong thrust on the use of virtualization technology to realize Infrastructure-as-a-Service (IaaS) has led enterprises to leverage subscription-oriented computing capabilities of public Clouds for hosting their application services. In parallel, research in academia has been investigating transversal aspects such as security, software frameworks, quality of service, and standardization. We believe that the complete realization of the Cloud computing vision will lead to the introduction of a virtual market where Cloud brokers, on behalf of end users, are in charge of selecting and composing the services advertised by different Cloud vendors. In order to make this happen, existing solutions and technologies have to be redesigned and extended from a market-oriented perspective and integrated together, giving rise to what we term Market-Oriented Cloud Computing.**

**In this paper, we will assess the current status of Cloud computing by providing a reference model, discuss the challenges that researchers and IT practitioners are facing and will encounter in the near future, and present the approach for solving them from the perspective of the Cloudbus toolkit, which comprises of a set of technologies geared towards the realization of Market Oriented Cloud Computing vision. We provide experimental results demonstrating market-oriented resource provisioning and brokering within a Cloud and across multiple distributed resources. We also include an application illustrating the hosting of ECG analysis as SaaS on Amazon IaaS (EC2 and S3) services.**

**KEY WORDS**: Cloud Computing, Platform-as-a-Service, Virtualization, Utility Computing, Market Oriented Computing

## 14.1. INTRODUCTION

In 1969, Leonard Kleinrock, one of the chief scientists of the original Advanced Research Projects Agency Network (ARPANET) project which seeded the Internet, said [26]: "As of now, computer networks are still in their infancy, but as they grow up and become sophisticated, we will probably see the spread of, "computer utilities", which, like present electric and telephone utilities, will service individual homes and offices across the country". This vision of computing utilities, based on a service-provisioning model, anticipated the massive transformation of the entire computing industry in the 21st century whereby computing services will be readily available on demand, like water, electricity, gas, and telephony services available in today's society. Similarly, computing service users (consumers) need to pay providers only when they access computing services, without the need to invest heavily or encounter difficulties in building and maintaining complex IT infrastructure by themselves. They access the services based on their requirements without regard to where the services are hosted. This model has been referred to as utility computing, or recently as Cloud computing [10].

Cloud computing delivers infrastructure, platform, and software (applications) as services, which are made available as subscription-based services in a pay-as-you-go model to consumers. In industry, these services are referred to as Infrastructure as a Service (IaaS), Platform as a Service (PaaS), and Software as a Service (SaaS), respectively. The Berkeley Report [3] released in Feb 2009 notes: "Cloud computing, the long-held dream of computing as a utility has the potential to transform a large part of the IT industry, making software even more attractive as a service".

Clouds aim to power the next generation data centers by architecting them as a network of virtual services (hardware, database, user-interface, application logic) so that users are able to access and deploy applications from anywhere in the world on demand at competitive costs depending on users Quality of Service (QoS) requirements [10]. It offers significant benefit to IT companies by freeing them from the low level tasks of setting up basic hardware (servers) and software infrastructures and thus enabling them to focus on innovation and creating business value for their services.

The business potential of Cloud computing is recognized by several market research firms including IDC (International Data Corporation), which reports that worldwide spending on Cloud services will grow from $16 billion by 2008 to $42 billion in 2012. Furthermore, many applications making use of Clouds emerge simply as catalysts or market makers that bring buyers and sellers together. This creates several trillion dollars of business opportunity to the utility/pervasive computing industry, as noted by Bill Joy, co-founder of Sun Microsystems [10].

Cloud computing has high potential to provide infrastructure, services and capabilities required for harnessing this business potential. In fact, it has been identified as one of the emerging technologies in IT as noted in "Gartner's IT Hype Cycle" (see Figure 14.1). A "Hype Cycle" is a way to represent the emergence, adoption, maturity and impact on applications of specific technologies.

Cloud computing is definitely at the top of the technology trend, reaching its peak of expectations in just 3-5 years. This trend is enforced by providers such as Amazon (http://aws.amazon.com), AT&T, Google, SalesForce (http://www.salesforce.com), IBM, Microsoft, and Sun Microsystems who have begun to establish new data centers for hosting Cloud computing applications such as social networks (e.g. Facebook- http://www.facebook.com, and MySpace- http://www.myspace.com), gaming portals (e.g. BigPoint- http://www.bigpoint.com), business applications (e.g., SalesForce.com), media content delivery, and scientific workflows. It is predicted that within the next 2-5 years, Cloud computing will become a part of mainstream computing; that is, it enters into the plateau of productivity phase.

Currently, the term Cloud computing mostly refers to virtual hosting solutions with some or no added value for customers. This market segment is known as Infrastructure-as-a-Service (IaaS) and concentrates the majority of the big companies operating in Cloud computing. The technology and the general concepts that characterize IaaS solutions are now largely developed and well established and many companies and users already adopt the Cloud option in order to save in infrastructure costs and access huge computing power on demand. The new challenges for what concerns the mainstream adoption of Cloud computing are more concentrated on how to make a profitable use of this technology and how to simplify the development of Cloud aware applications. In particular there is an entire market related to the delivery of platforms and tools for building applications that are hosted in the Cloud or leverage Cloud services for many of their tasks. In this sense, the Cloudbus Toolkit for Market Oriented Cloud Computing provides a set of tools and technologies that, taken together, contribute to realize the vision of Cloud computing. It approaches this challenge from a market-oriented perspective, which is one of the driving factors of this technology.

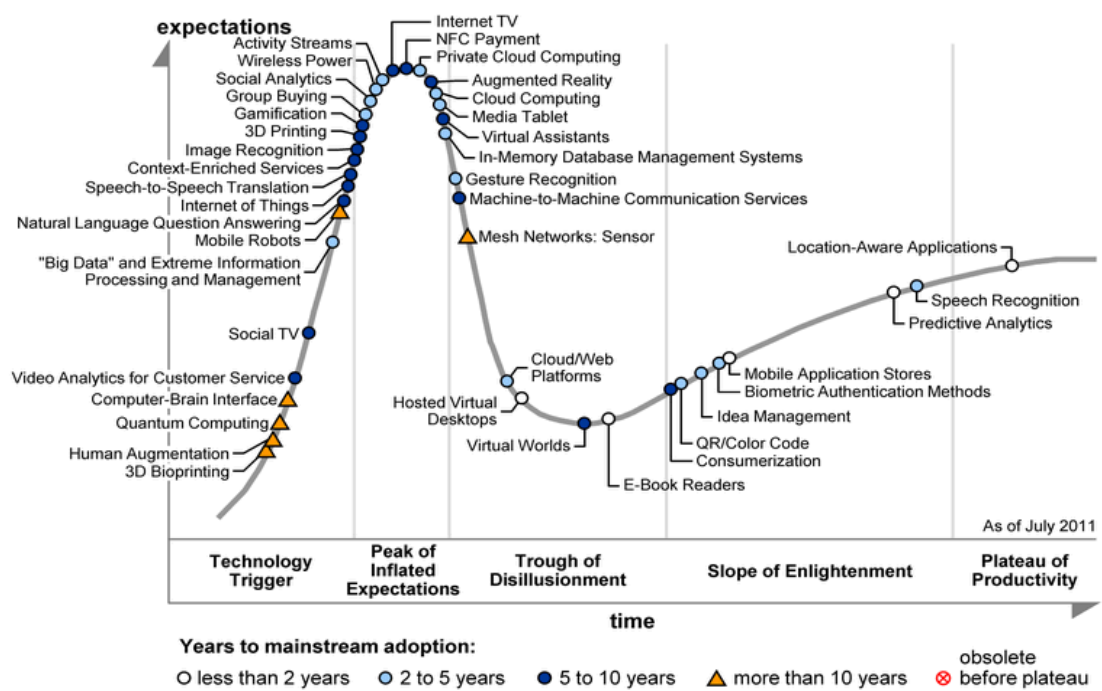

Figure 14.1: Gartner 2011 Hype Cycle of Emerging Technologies.

The rest of the paper is organized as follows: Section 14.2 presents a high-level definition of Cloud computing followed by open challenges and a reference model; Section 14.3 presents Cloudbus vision and architecture in conformance with the high-level definition; Section 14.4 lists specific technologies of the Cloudbus toolkit that have made the vision a reality; Section 14.5 discusses the integration of the Cloudbus toolkit with other Cloud management Technologies; and Section 14.6 presents experimental results demonstrating market-oriented resource provisioning within a Cloud and across distributed resources along with hosting of ECG analysis as SaaS on Amazon IaaS (EC2 and S3) services. Finally, Section 14.7 concludes the paper providing insights into future trends in Cloud computing.

## 14.2. CLOUD COMPUTING

Cloud computing [3, 10] is an emerging paradigm that aims at delivering hardware infrastructure and software applications as services, which users can consume on a pay-per- use-basis. As depicted in Figure 14.1, Cloud computing is now at the peak of its hype cycle and there are a lot of expectations from this technology. In order to fully understand its potential, we first provide a more precise definition of the term, introduce a reference model for Cloud computing, provide a brief review of the state of the art, and briefly sketch the challenges that lie ahead.

**14.2.1 Cloud Definition and Market-Oriented Computing**

Due to rapid advances in Cloud computing paradigm, it means different things to different people. As a result, there are several definitions and proposals [42]. Vaquero et al. [42] have proposed a definition that is centered on scalability, pay-per-use utility model and virtualization. According to Gartner, Cloud computing is a style of computing where service is provided across the Internet using different models and layers of abstraction. The cloud symbol traditionally represents the Internet. Hence, Cloud computing refers to the practice of moving computing to the Internet. Armbrust et al. [3] observe that "Cloud computing refers to both the applications delivered as services over the Internet and the hardware and system software in the data centers that provide those services". This definition captures the real essence of this new trend, where both software applications and hardware infrastructures are moved from private environment to third parties data centers and made accessible through the Internet. Buyya et al. [10] define a Cloud as a type of parallel and distributed system consisting of a collection of interconnected and virtualized computers that are dynamically provisioned and presented as one or more unified computing resources based on service-level agreements". This definition puts Cloud computing into a market oriented perspective and stresses the economic nature of this phenomenon.

The key feature, emerging from the above characterizations is the ability to deliver both infrastructure and software as services that are consumed on a pay-per-use-basis. Previous trends were limited to a specific class of users, or specific kinds of IT resources; the approach of Cloud computing is global and encompasses the entire computing stack. It provides services to the mass, ranging from the end-users hosting their personal documents on the Internet to enterprises outsourcing their entire IT infrastructure to external data centers. Service Level Agreements (SLAs), which include QoS requirements, are set up between customers and Cloud providers. An SLA specifies the details of the service to be provided in terms of metrics agreed upon by all parties, and penalties for violating the expectations. SLAs act as a warranty for users, who can more comfortably move their business to the Cloud. As a result, enterprises can cut down maintenance and administrative costs by renting their IT infrastructure from Cloud vendors. Similarly, end-users leverage the Cloud not only for accessing their personal data from everywhere, but also for carrying out activities without buying expensive software and hardware.

Figure 14.2 shows the high level components of the service-oriented architectural framework consisting of clients brokering and coordinator services supporting utility-driven management of Clouds: application scheduling, resource allocation and migration of workloads. The architecture cohesively couples the administratively and topologically distributed storage and compute capabilities of Clouds as parts of a single resource leasing abstraction [10]. The system will ease the cross-domain integration of capabilities for on-demand, flexible, energy-efficient, and reliable access to the infrastructure based on emerging virtualization technologies [1, 4].

Market oriented computing in industry is getting real as evidenced by developments from companies such as Amazon. For example, EC2 started with flat pricing then moved to pricing based on service difference and recently introduced auction based models, using spot instances (http://aws.amazon.com/ec2/spot-instances).

The Cloud Exchange (CEx) acts as a market maker for bringing together service producers and consumers. It aggregates the infrastructure demands from the application brokers and evaluates them against the available supply currently published by the Cloud Coordinators. It aims to support trading of Cloud services based on competitive economic models such as commodity markets and auctions. CEx allows the participants (Cloud Coordinators and Cloud Brokers) to locate providers and consumers with fitting offers. Such markets enable services to be commoditized and thus, can pave the way for the creation of dynamic market infrastructure for trading based on SLAs. The availability of a banking system within the market ensures that financial transactions pertaining to SLAs between participants are carried out in a secure and dependable environment. Every client in the Cloud platform will need to instantiate a Cloud brokering service that can dynamically establish service contracts with Cloud Coordinators via the trading functions exposed by the Cloud Exchange.

This is a broad vision about how a future Market-Oriented Cloud Computing system should be structured. The available technologies for Cloud computing are components that can be used to realize this vision. Before exploring them, we will introduce a Cloud computing reference model provides an organic view of a Cloud computing system and will be used to classify the state of the art.

### 14.2.2 Cloud Computing Reference Model

Figure 14.2 provides a broad overview of the scenario envisioned by Cloud computing. This scenario identifies a reference model into which all the key components are organized and classified. As previously introduced, the novelty of this approach encompasses the entire computing stack: from the system level, where IT infrastructure is delivered on demand, to the user level, where applications transparently hosted in the Cloud are accessible from anywhere. This is the revolutionary aspect of Cloud computing that makes service providers, enterprises, and users completely rethink their experience with IT.

The lowest level of the stack is characterized by the physical resources, which constitute the foundations of the Cloud. These resources can be of different nature: clusters, data centers, and desktop computers. On top of these, the IT infrastructure is deployed and managed. Commercial Cloud deployments are more likely to be constituted by data centers hosting hundreds or thousands of machines, while private Clouds can provide a more heterogeneous environment, in which even the idle CPU cycles of desktop computers are used to handle the compute workload. This level provides the "horse power" of the Cloud.

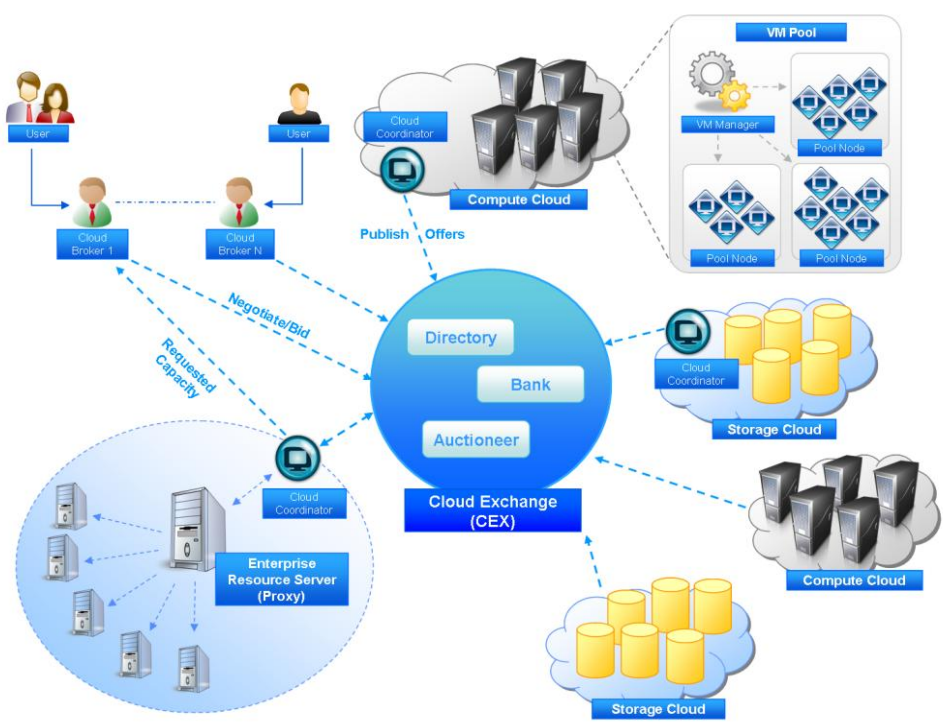

**Figure 14.2: Utility-oriented Clouds and their federated network mediated by Cloud exchange.**

The physical infrastructure is managed by the core middleware whose objectives are to provide an appropriate runtime environment for applications and to utilize the physical resources at best. Virtualization technologies provide features such as application isolation, quality of service, and sand boxing. Among the different solutions for virtualization, hardware level virtualization and programming language level virtualization are the most popular. Hardware level virtualization guarantees complete isolation of applications and a fine partitioning of the physical resources, such as memory and CPU, by means of virtual machines. Programming level virtualization provides sand boxing and managed executions for applications developed with a specific technology or programming language (i.e. Java, .NET, and Python). Virtualization technologies help in creating an environment in which professional and commercial services are integrated. These include: negotiation of the quality of service, admission control, execution management and monitoring, accounting, and billing.

Physical infrastructure and core middleware represent the platform where applications are deployed. This platform is made available through a user level middleware, which provides environments and tools simplifying the development and the deployment of applications in the Cloud. They are: web 2.0 interfaces, command line tools, libraries, and programming languages. The user-level middleware constitutes the access point of applications to the Cloud.

At the top level, different types of applications take advantage of the offerings provided by the Cloud computing reference model. Independent software vendors (ISV) can rely on the Cloud to manage new applications and services. Enterprises can leverage the Cloud for providing services to their customers. Other opportunities can be found in the education sector, social computing, scientific computing, and Content Delivery Networks (CDNs).

### 14.2.3 State of the Art in Cloud Computing

It is quite uncommon for a single solution to encompass all the services described in the reference model. More likely, different vendors focus on providing a subclass of services addressing a needs of a specific market sector while research projects are more interested in facing the challenges of a specific aspect of Cloud computing, such as scheduling, security, privacy, and virtualization. In this section, we will review the research works and the most prominent commercial solutions for delivering Cloud computing based software systems. By following the previous reference model it is possible to classify the available options into three main categories: Software-as-a-Service (SaaS), Platform-as-a-Service (PaaS), and Infrastructure / Hardware-as-a-Service (IaaS/HaaS), as depicted in Figure 14.3. Table 14.1 summarizes the main characteristics of these categories and provides some examples of organizations offering respective services.

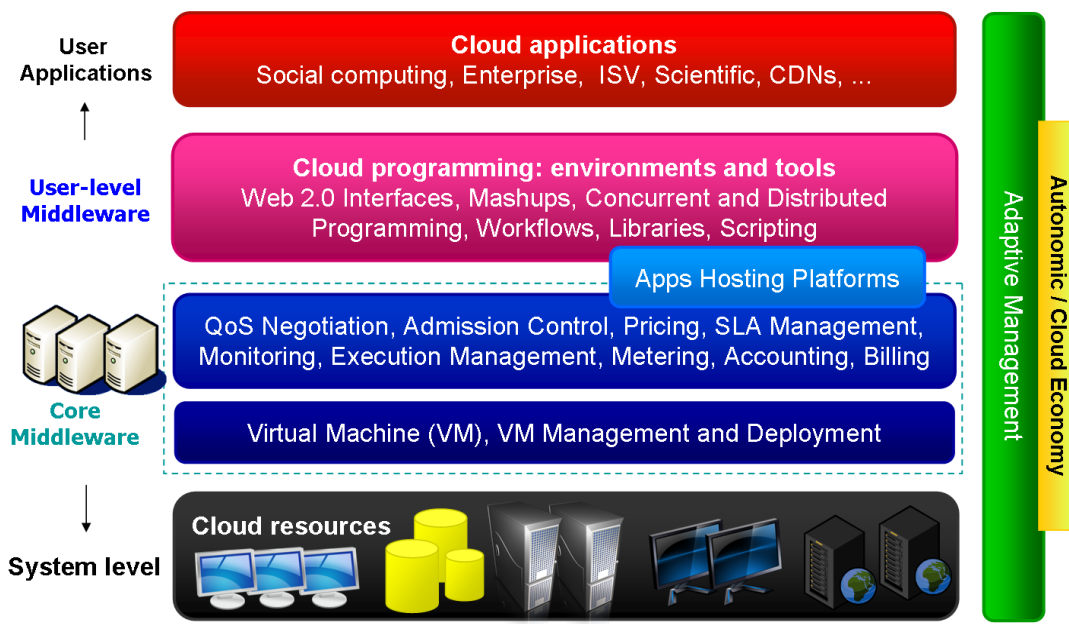

**Figure 14.3: Cloud computing reference model.**

### 14.2.3.1 Infrastructure as a Service

Infrastructure-as-Service (IaaS) or Hardware-as-a-Service (HaaS) solutions deliver IT infrastructure based on virtual or physical resources as a commodity to customers. These resources meet the end user requirements in terms of memory, CPU type and power, storage, and, in most of the cases, operating system as well. It is possible to identify two different approaches: pure IaaS solutions that provide both a management infrastructure and the physical hardware where the infrastructure is deployed, and IaaS implementations that concentrate only on providing a management infrastructure and are meant to be deployed on top of a physical existing infrastructure provided by the user.

The idea of using hardware virtualization technologies for providing executing environments on demand is not new. The first attempts to provide a virtual machine based execution environment for applications can be found in the Denali project [47]. The focus of Denali was to provide a scalable infrastructure able to support the management of a large number of server applications by using lightweight virtual machines. Figuereido et al. [17] investigated the use of virtual machine images for customizing the execution environments in Grids. VMPlant [27] is a framework that embodies these concepts and provides a management infrastructure of virtual machine within a computing Grid. On the same line, Virtual Workspaces [22] provide configurable execution environments that are dynamically deployed by means of virtual machine images in a Grid infrastructure. An evolution of these concepts is Nimbus [23], which constitutes a complete realization of the IaaS model for science clouds. It is a set of open source tools, when puts together, contribute to deliver an Infrastructure-as-a-Service solution mostly focused on scientific applications. OpenNebula [41] and Eucalyptus [33] constitute a complete platform for delivering IaaS solutions. Eucalyptus is an open-source framework that turns a collection of clusters into a computing and storage cloud. It provides interface compatibility with, and constitutes an alternative to, Amazon EC2 and S3 allowing users to build an Amazon-like private cloud and to migrate naturally to the public infrastructure later. OpenNebula is a virtual machine manager that can be used to deploy virtualized services on both a local pool of resources and on external IaaS clouds. Together with Haizea [41], a resource lease manager that can act as a scheduling backend for OpenNebula, it provides advanced features such as resource reservation and preemption. OpenNebula and Haizea have been developed under the RESERVOIR project [39] aiming at defining advanced system and service management approach that will serve as infrastructure for cloud computing implementations.

Pure IaaS solutions are more likely to be found in industry: Amazon is one of the major players in this field. Amazon Elastic Compute Cloud (EC2) provides a large computing infrastructure and a service based on hardware virtualization. By using Amazon Web Services, users can create Amazon Machine Images (AMIs) and save them as templates from which multiple instances can be run. It is possible to run either Windows or Linux virtual machines, for which the user is charged per hour for each of the instances running. Amazon also provides storage services with the Amazon Simple Storage Service (S3), users can use Amazon S3 to host large amount of data accessible from anywhere. Joyent provides customers with infrastructure, hosting, and application services. It has been particularly successful in scaling collaborative applications such as Linkedin (http://www.linkedin.com) and Facebook. Facebook, for example, now hosts nearly 300 million of users that are seamlessly using Amazon Cloud services. Other relevant implementations of pure IaaS solutions are GoGrid, ElasticHosts, Rackspace, Flexiscale. Some vendors are mostly focused on providing a software management infrastructure that allows users to exploit at best existing virtual infrastructure. Commercial solutions of this kind rely on existing pure IaaS vendors and provide added value on top of them. RightScale provides a management layer aiming to eliminate the vendor lock- in by letting the user to choose the specific virtual infrastructure (Amazon, VMWare, etc) and software stack to compose for their virtual environment (SkyTap). Other vendors, such as CloudCentral and Rejila, add specific features such as facilities for composing your own virtual infrastructure and or automated application packaging and deployment. Other solutions are completely specialized in providing a flexible and full featured virtual infrastructure design environment and do not provide bare metal virtual servers or storage (Elastra, CohesiveFT).

Table 14.1: Cloud computing services classification.

| Category | Characteristics | Product Type | Vendors & Products |
|---|---|---|---|
| SaaS | Customers are provided with applications that are accessible anytime and from anywhere | Web applications and services (Web 2.0) | SalesForce.com (CRM), Clarizen.com (Project Management), Google Mail (Automation) |
| PaaS | Customers are provided with a platform for developing applications hosted on the Cloud | Programming APIs and frameworks; Deployment system. | Google AppEngine, Microsoft Azure, Manjrasoft Aneka |
| IaaS/HaaS | Customers are provided with virtualized hardware and storage on top of which they can build their infrastructure | Virtual machines management infrastructure, storage management | Amazon EC2 and S3; GoGrid; Nirvanix |

**14.2.3.2 Platform as a Service**

Platform as a Service solutions provide an application or development platform in which users can create their own application that will run on the Cloud. More precisely, they provide an application framework and a set of API that can be used by developers to program or compose applications for the Cloud. Currently, most of the research and the industrial effort have been put into providing IaaS solutions, which are commonly identified as Cloud computing. PaaS solutions are more likely to be explored in the next coming years, once the technologies and the concepts of infrastructure provisioning are fully established. For this reason there are a limited number of implementations for this approach in both the academy and the industry. We can categorize the PaaS approach into two major streams: those who integrate an IT infrastructure on top of which applications will be executed as a part of the value offering and those, which do not. Solutions that include an IT infrastructure are most likely to be found in the industry, while the other ones are more common in the academy.

MapReduce [14] has gained a considerable success as a programming model for the Cloud. Google has proposed it for processing massive quantities of data on large-scale distributed infrastructures. It is characterized by programming model expressing distributed applications in terms of two computations, map and reduce, and a fault tolerant distributed file system that is optimized for moving large quantities of data. Hadoop [48] is an open source implementation of MapReduce and has been utilized as Cloud programming platform on top of the Amazon EC2 (Elastic MapReduce) and the Yahoo Cloud Supercomputing Cluster. Other research works and commercial implementations adopting the PaaS approach are mostly focused on providing a scalable infrastructure for developing web applications. AppEngine (http://code.google.com/appengine) is Platform-as-a-Service solution proposed by Google for developing scalable web applications executed on its the large server infrastructure. It defines an application model and provides a set of APIs that allow developers to take advantage of additional services such as Mail, Datastore, Memcache, and others. Developers can develop their application in different languages (Python, Java, and other JVM based languages), upload it to AppEngine that will execute it in a sandboxed environment and automatically scale up and down. AppScale [11] is an open source implementation of AppEngine, developed at the University of California, Santa Barbara. It enables the execution of AppEngine applications on local clusters and can utilize Amazon EC2 or Eucalyptus based clouds to scale out applications. It is meant to provide a framework for scientists to do research on programming cloud applications. Heroku (http://www.heroku.com) is a Cloud computing platform that automatically scales web applications based on Ruby on Rails (http://rubyonrails.org). Very few implementations propose a platform for developing any kind of application in the Cloud. Azure (http://www.microsoft.com/windowsazure) is a cloud service operating system that serves as the development, run-time, and control environment for the Azure Services Platform. By using the Microsoft Azure SDK, developers can create services that leverage the .NET Framework. These services have to be uploaded through the Microsoft Azure portal in order to be executed on top of Windows Azure. Additional services, such as workflow execution and management, web services orchestration, and access to SQL data stores, are provided to build enterprise applications. Extreme Application Platform (XAP) (http://www.gigaspaces.com/xap) commercialized by GigaSpaces, is a middleware for developing ultra-fast, scalable, distributed applications. It is based on the concept of space; represent a shared environment that can be used as a fast in memory distributed store, execution runtime for applications, and message bus. By using XAP it is possible to define policies for elastically scale applications according to their needs. SaaSGrid (http://apprenda.com) commercialized by Apprenda is a software development platform specifically designed for developing SaaS applications. The Granules [2] project is a lightweight streaming-based runtime for cloud computing. It orchestrates the concurrent execution of applications on multiple machines. The runtime manages an applications execution through various stages of its lifecycle: deployment, initialization, execution and termination. Force.com (http://www.salesforce.com/platform) and CloudHarbor.com (http://www.cloudharbor.com) are similar examples but mostly focused on the development of Business Process Modeling (BPM) applications.

As part of the Cloudbus Toolkit, Aneka [43] is a pure PaaS implementation commercialized by Manjrasoft, is a pure PaaS implementation for developing scalable applications for the Cloud. The core value of Aneka is a service oriented runtime environment that is deployed on both physical and virtual infrastructures and allows the execution of applications developed by means of various programming models. More details will be given in section 14.4.1.

**14.2.3.3  Software as a Service**

Software as a Service solutions are at the top end of the Cloud computing stack and they provide end users with an integrated service comprising hardware, development platforms, and applications. Users are not allowed to customize the service but get access to a specific application hosted in the Cloud. The SaaS approach [5] for delivering IT services is not new but it has been profitably integrated into the Cloud computing stack by providing an on-demand solution for software applications. Examples of SaaS implementations are the services provided by Google for office automation, such as Google Mail, Google Documents, and Google Calendar, which are delivered for free to the Internet users and charged for professional quality services. Examples of commercial solutions are SalesForce.com (http://www.salesforce.com) and Clarizen.com (http://www.clarizen.com), which provide online CRM (Customer Relationship Management) and project management services, respectively. Appirio (http://www.appirio.com) is an integrated solution that provides complete support for any management aspect of modern enterprises from project management to resource planning. The peculiarity of Appirio is the ability of integrating into the platform additional services exposed by other Clouds such as Amazon EC2, SalesForce.com, Google AppEngine, and Facebook.

### 14.2.3.4 Alliances and Standardization Initiatives

Research and activities on Cloud computing have also investigated other aspects, which are transversal to the classification previously introduced. These aspects include: security, privacy, standardization, and interoperation.

Security and privacy are one of the major research areas in Cloud computing besides the development of frameworks. In particular, trust has been reported to be one of the most important issues when considering moving to the Cloud. On this topic Li and Ping [30] developed a trust model for the enhancing security and interoperation among Clouds. Pearson et al. investigated the management from the perspective of Cloud Services design [36] and of data encryption [38], while other research works focused on access control to the Cloud and identity management [20, 29, 37]. Security is not of interest only in academia but a lot of IT practitioners face and discuss the challenges of security in the Cloud. As an example, Cloud Security Alliance is an initiative whose mission is to promote the use of best practices for providing security assurance within Cloud computing, and provide education on the uses of Cloud computing to help secure all other forms of computing.

Standardization and interoperability is another important area of research in Cloud computing. Currently only few works have been investigated these topics and the most relevant outcomes are the Open Cloud Manifesto (http://www.opencloudmanifesto.org) and the Open Virtualization Format (OVF)[1]. The Open Cloud Manifesto represents an initiative, supported by the major players in Cloud computing, for the promotion of Open Clouds characterized by interoperability between providers and true scalability for applications. The Open Virtualization Format is an open standard for packaging and distributing virtual appliances or more generally software to be run in virtual machines. These initiatives are still at an early stage and more research has to be pursued in this field.

### 14.2.4 Open Challenges

Cloud computing introduces many challenges for system and application developers, engineers, system administrators, and service providers [3, 10, 12]. Figure 14.4 provides an overview of the key challenges. In this section, we will discuss some of them.

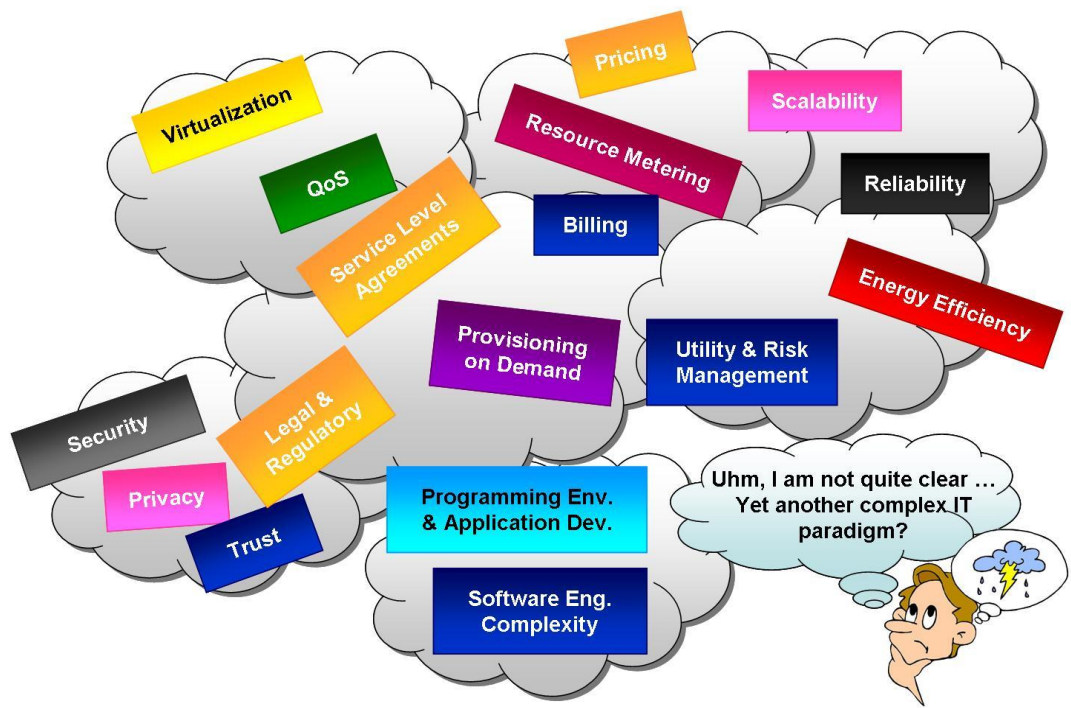

Figure 14.4: Open challenges in Cloud computing.

#### 14.2.4.1 Virtualization

Virtualization enables consolidation of servers for hosting one or more services on independent virtual machines in a multi-tenancy manner. When a large number of VMs are created they need to be effectively managed to ensure that services are able to deliver quality expectations of users. That means, VMs need to be migrated to suitable servers to ensure that agreed QoS is delivered; and consolidate them later to a fewer number of physical servers when the demand decreases. These capabilities draw challenging questions:

---

[1] http://www.dmtf.org/standards/published documents/DSP0243 1.0.0.pdf

- How does the service provider guarantee scalability to its users?
- How to make a profitable use of virtualization technology to satisfy customer requirements and infrastructure capabilities?

It is a customary practice not to disclose the amount of compute/storage resources a service provider has to its customers. On this setting, a customer may choose a particular provider solely based on its reputation and advertised capabilities. When the service provider gets a large number of requests, it may have to overload its hardware to fulfill these requests. The challenge here is the capability to manage a sheer number of requests for VMs and the load on the infrastructure [40]. Even though theoretically it may be possible to scale out, replicate VMs [28], practically any service provider may limit its resources for several reasons: managerial, cost, risk, etc. There are software and hardware barriers when trying to instantiate large number of VMs in a data center [21]. In such cases, when service brokers overprovision resources across datacenters with an aim to accommodate large number of user requests, virtualization limitation may result in violation of contracts. This is always a challenge when trying to balance over provisioning using virtualization techniques.

A service provider has to adopt a technology that suits its customer needs as well as matches its infrastructure capabilities. A perfect matching has a long-term effect on revenues, market impact and sustainability. For e.g. different Cloud vendors may choose to adopt any of the hypervisors (e.g. Xen, Citrix, Hyper-V, VMware, etc.) as per their service characteristics and requirements. This brings a divide between the interoperability of VM images between Cloud providers. The co-existence of competitive technologies tends to lower the effectiveness of each of them, until one becomes a common standard.

**14.2.4.2  Security, Privacy and Trust**

One of the major concerns when moving to Clouds is related to security, privacy, and trust. Security in particular, affects the entire cloud computing stack [36]. The Cloud computing model promotes massive use of third party services and infrastructures to host important data or to perform critical operations. In this scenario, the trust towards providers is fundamental to ensure the desired level of privacy for applications hosted in the Cloud. The obvious questions are:

- How to secure the data and computation on the VMs managed by a Cloud service provider?
- What is the role of a VM management software in ensuring security from both providers and users point of view and isolation from application and users data? In particular, what are the restrictions on obtaining and using statistical data out of hosted services?

- How to manage access to VM images, track the provenance of images, and provide users and administrators with efficient image filters and scanners that detect and repair security violations? [46]
- How do we trust third party software that is a part of the Cloud infrastructure? What are the boundaries/restrictions for engaging third parties in the service provider management chain?
- What are the standards of security, privacy and trust in computation, data and identity of Cloud users [36]?

One of the major concerns for the end users of Cloud computing services is the risk of leakage of data deployed to Cloud computing services. A virtual machine manager/resource allocator manages the VM nodes in a data center. As these virtual nodes are deployed on top of physical hardware, there is always a super user (privileged user) from the provider's side who has access to the VM state and the physical node. Any accidental or intentional access/leak of data processed by the VMs cannot be completely ruled out. Even encrypting data would not be of much help as the raw data are processed in the memory. Both data and computations are susceptible to attacks resulting from any intruder's VM inspection, unauthorized VM migrations to any physical nodes. The other side of this problem can be analyzed from a provider side. Currently, any Cloud user can use any Software on the VM as long as the user pays for the usage of the services. This user might be running a spamming network/software in the Cloud. Cloud service providers face an unabated challenge to identify and restrict malicious attempts by users of its services. This issue defines a new boundary on the capabilities of the VM management software. If the underlying hypervisor is allowed to transparently monitor the processes a VM is running, the use of malicious software could be restricted. In this process, a service provider may choose to offload part of its responsibilities (monitoring, identifying and accounting) to third party application vendors. In such cases, customer privacy is directly or indirectly affected by the functionality and terms of operation of those tertiary units.

Cloud service providers expose operating systems, applications and utilities as images for public as well as private use. A user can lease these images to instantiate a VM or use the application/s bundled in the instantiated image. Users may customize the image and then store them for future usage. These public/private images are shared images with access rights managed by the service provider upon the user's request. The responsibility of checking the integrity of these images in terms of security risks to other VMs running in the data center lies in the service provider's side. It is a challenge to continuously maintain provenance of images, their composition and access rights in a large public Cloud computing infrastructure.

Trusting a Cloud service provider to secure user data, computation and the compliance terms laid out in the SLA is now a matter of innovation. The tools and capabilities provided by providers to monitor QoS satisfactions would need to be audited by a third party that both provider and end user trust. Organizing such a trust network in Cloud computing by not compromising its utility, flexibility and economy is a challenge.

The lack of and insufficiency of standards in maintaining privacy of computation, data and identity of end users elevates the challenges in using Cloud computing services. At present, traditional tools and models used to enforce a secure and reliable environment from a security point of view are the only ones available. As previously discussed in section 14.2.3.4 this area is very interesting from a research point of view, and some early works have already been done. These could be used as a starting point for building the security infrastructure of the Cloud for the future.

**14.2.4.3  Legal and Regulatory**

Besides security, there are legal and regulatory issues that need to be taken care of. Cloud service providers may choose to host user application data anywhere on the planet. The physical location of data centers and clusters determines the set of laws that can be applied to the management of data. For example, specific cryptography techniques could not be used because they are not allowed in some countries. Simply, specific classes of users, such as banks, would not be comfortable to put their sensitive data into the Cloud, in order to protect their customers and their business. At present, a conservative approach is taken for what concerns hosting sensitive data. An interesting initiative is the concept of availability zones promoted by Amazon EC2. Availability zones identify a set of resources that have a specific geographic location. Currently there are two regions grouping the availability zones: US and Europe. Although this initiative is mostly concerned with providing of better services in terms of isolation from failures, network latency, and service down-time, it could be an interesting example for exploring legal and regulatory issues.

**14.2.4.4  Service Level Agreements and Quality of Service**

Service Level Agreements define the functional and non-functional characteristics of Cloud services that is agreed by both the customer and the provider. The common parameters that define a SLA are: pricing model, usage model, resource metering, billing, and monitoring. In most cases, the desired level of security is also established within a SLA. When a service provider is unable to meet the terms stated in the SLA, a violation occurs. For example, an IaaS Cloud service provider may guarantee a minimum response time from a VM, minimum storage space, reliability of data, etc. However, if a customer does not get the desired response time, runs out of virtual disk space or is met with frequent errors, the SLA is violated. The SLA also defines a penalty model to compensate the customer in case of violations. At present, the adopted solution for pricing falls into the pay-as-you-go model and the users are charged according to the usage of the Cloud services. With constant changing of customer requirements, providers face the following challenges:

- How to guarantee QoS satisfactions and prevent SLA violations?
- How to manage Cloud services to meet the SLA terms for increasing customers and for their ever-increasing demands?
- How to manage SLA in a Cloud computing environment?

The notion of QoS satisfaction varies across customers as every user has its own requirements. Some general metrics from users prospective are: amount of aggregate CPU power for the VMs instantiated, minimum bandwidth available, number and size of input/output devices (e.g. storage volumes, virtual hardware, etc.), average response time, etc. Typically, a customer is more inclined to request a statistical bound on most of these parameters than an average [50]. At the moment, no Cloud service providers are guaranteeing the minimum QoS for any of these metrics. From a provider's point of view, it still remains a challenge to provision, manage and predict the use of its Cloud services in the long run. That difficulty obstructs it to state concrete SLA terms in writing with its customers. With the increasing number of users, most violations are likely to happen during load fluctuations due to the lack of either sufficient resources or weakness in managing VMs at the provider's side. In this direction, Patel et al. [35] have proposed a mechanism for managing SLAs in a Cloud computing environment using the Web Service Level Framework (WSLF) [24]. They propose using dynamic schedulers for measuring parameters, enabling measurements through third parties, and modeling penalties as financial compensations (moderated via a third party), to adapt web SLA to a Cloud environment.

More sophisticated and flexible pricing policies that take into account SLA violations have to be developed and put in place in order to devise an efficient pricing model for the Cloud computing scenario. As services are offered on a subscription basis, they need to be priced based on users QoS expectations that vary from time to time. The complexity of enabling a SLA is higher in a multi-tenancy environment [49], where many businesses (i.e. tenants) have varying QoS requirements. It is also important to ensure that whenever service providers are unable to meet all SLAs, their violation needs to be rectified so that customers do not have to bear the loss resulting from service provider's incompetence.

**14.2.4.5  Energy Efficiency**

Data centers are expensive to operate as they consume huge amount of power [31]. The combined energy consumption of all data centers worldwide is equivalent to the power consumption of Czech Republic. As a result, their carbon footprint on the environment is rapidly increasing. In order to address these issues, energy efficient resource allocation and algorithms need to be developed. The challenges are as follows:

- How to balance energy consumption and optimal performance of data centers so that users can be charged at a nominal rate?
- How to choose locality of data centers so that data security, operation cost, and energy consumption meet the terms in the SLA signed with users?

The performance of data centers depends on the provisioning and usage of its hardware devices by the VM management software depending on user needs. As more CPUs are used, the temperature of the hardware increases. This requires cooling of the data center. Hence, performance of the data center and energy consumption is directly related to each other. For every increase in energy consumed, the cost of operation of the data center adds up. This cost may get transferred to the users unless the provider balances the performance and energy consumption. Placing the data centers in cold regions such as Iceland is seen to be a viable option. However, there are concerns about the locality of data as users may restrict where their data is placed. Placement and sizing of data centers presents a challenging optimization problem, involving several factors [19].

**14.2.4.6  Programming Environments and Application Development**

Cloud computing introduces practical and engineering problems to solve. Cloud computing infrastructures need to be scalable and reliable. In order to support a large number of application service consumers from around the world, Cloud infrastructure providers (i.e., IaaS providers) have been establishing data centers in multiple geographical locations to provide redundancy, ease of access, and ensure reliability [15]. Cloud environments need to provide seamless/automatic mechanisms for scaling their hosted services across multiple, geographically distributed data centers in order to meet QoS expectations of users from different locations. The scaling of applications across multiple-vendor infrastructures requires protocols and mechanisms needed for the creation of inter-cloud environments.

From the perspective of applications the development of platform and services that take full advantage of the Cloud computing model, constitute an interesting software engineering problem. The immediate challenges are:

- Should the application logic and its scalability be handled by the application itself or be

- entrusted to a third party middleware?
- How to provide application developers the technical know-how and the intricacies of multiple data centers, platforms and services?
- How to define the terms and conditions for licensing the usage and interoperability between numerous SaaS in Clouds?

Numerous middleware are being designed to handle the scalability of applications so that application designers are isolated from the intricacies of Cloud platforms. However, this practice results in: a) applications having to rely on a generic middleware to scale their logic, b) developers are usually restricted in following a confined set of APIs to use the middleware, which instead limits the features of applications. If application developers were to know the details of the data center they would deploy their application in, it would make the application custom designed for high performance in specific data centers. This duality is a challenge as Cloud providers would not want to disclose their hardware details and application developers are limited using APIs from third party middleware. For SaaS providers, licensing has become a major issue. Moving an application to a public Cloud would make proprietary software accessible to millions. This is a challenge as software vendors are wondering how to respect the boundaries of Open Source technologies and licensed software in Clouds, yet making them interoperable.

### 14.2.4.7 Applications on Clouds

At present there are numerous real-world applications that are running on distributed clusters around the world. However, only a few of them would be able to utilize Cloud resources with minor modifications. This is due to the fact that legacy applications were designed to operate on physical hardware with heavy optimizations targeting storage, input/output, communication etc. Cloud computing offers a different paradigm where traditional assumptions on hardware devices and software models may not always work. Input/output throughput, for example, may be different depending upon the location of the VM instance allocated for an application and the storage hardware used. Similarly, other attributes of an application such as: user experience, distribution, maintenance have new issues when applications are moved to Clouds. The questions that are important to ask before moving applications to Clouds are:

- How to map application attributes to Cloud attributes [13]?
- Are all applications "Cloud ready"?
- Should an application be using multiple Cloud services to rely on or a single Cloud service provider?

Application attributes, such as data requirements, platform, communication, distribution, security etc., may be related to different layers of the Cloud stack. It may satisfy the requirements by combining services from different Cloud vendors. But, combining different service providers brings along higher cost, risks, managerial difficulties and interoperation issues. Applications need to be "Cloud ready" before they can reap the benefit of what Cloud computing has to provide.

**14.2.4.8  Standardization**

As Cloud is becoming a commonly used environment for hosting applications, numerous tools and services are available for use from each vendor. Due to a lack of standardization, these tools are not fully compatible with each other. This only accelerates the divide between Cloud service providers, limiting the interoperability between services hosted by each provider. For example, an application may need to implement ad-hoc connectors to utilize IaaS solutions from different vendors.

As already introduced in section 14.2.3.4, an effort towards standardization is the Open Virtualization Format (OVF), an open standard for packaging and distributing virtual appliances or more generally software to be run in virtual machines. Although major representative companies in the field (Microsoft, IBM, Dell, HP, VMware, and XenSource) are supporting the initiative, which resulted into a preliminary standard by the Distributed Management Task Force, the OVF specification only captures the static representation of a virtual instance. Hence, it is mostly used as canonical way of distributing virtual machine images. Many vendors and implementations simply use OVF as an import format and convert it into their specific runtime format when running the image. In the management layer of the Cloud computing stack, the OGF Open Cloud Computing Interface (OCCI) is working towards specifications for remote management of Cloud computing services. Their specification would help standardize the development of tools that govern the functionality of deployment, scaling and monitoring of VMs and/or workloads running as part of elastic Cloud services.

Standardizing every aspect of IaaS, PaaS or SaaS is challenging. Vendors try to make their products different than their competitors to gain a better market share. Having a unique and highly regarded capability usually draws lots of customers initially. However, after many companies evolve and the capability is common among all vendors, it becomes a standard. But until that happens, users may not be able to utilize existing capabilities across all Cloud service providers. Standardizing on a single cloud provider can lead to data lock-in or application architecture or application development lock-in [32]. Additional effort has to be put to define and enforce standards for both customer and service provider's satisfaction.

These are some of the key challenges that need to be addressed for a successful adoption of the Cloud computing paradigm into the mainstream IT industry. R&D initiatives in both academia and industry are playing an important role in addressing these challenges. In particular, the outcome of such research in terms of models, software frameworks, and applications constitute the first tools that can be used to experience Cloud computing. The Cloudbus Toolkit is a step towards this goal.

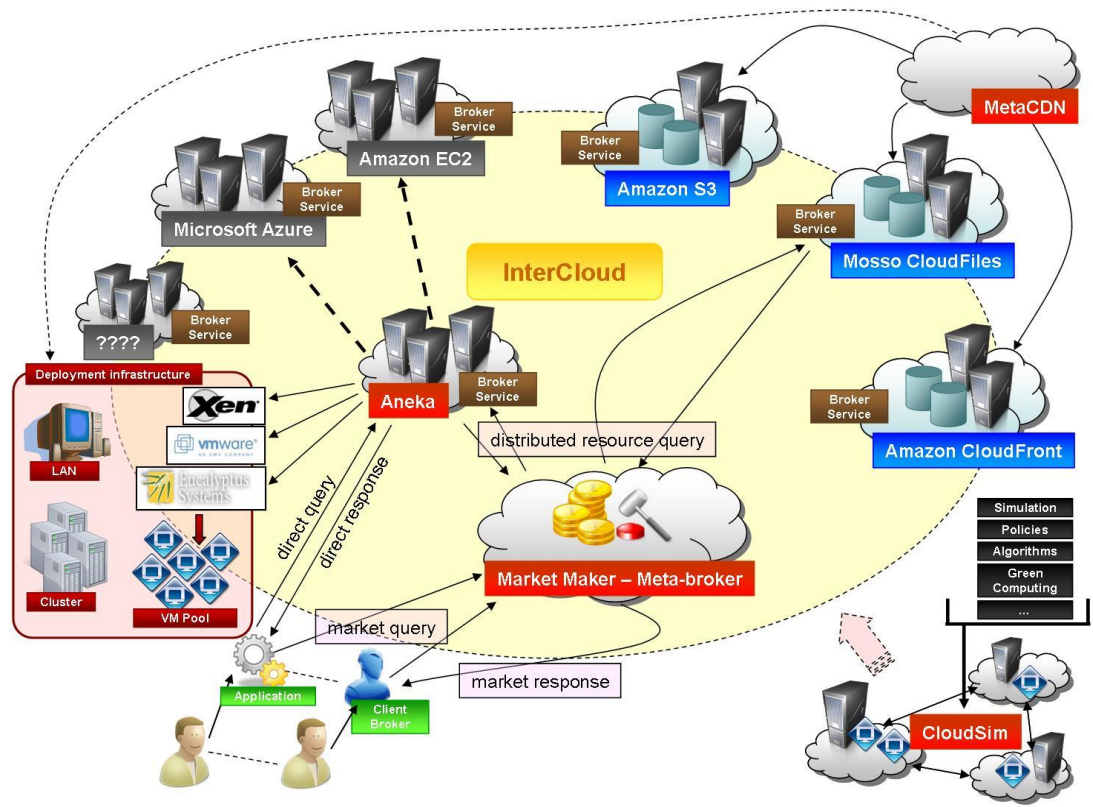

**Figure 14.5: Cloud computing marketplace.**

## 14.3. CLOUDBUS: VISION AND ARCHITECTURE

Figure 14.5 provides a glimpse in the future of Cloud computing. A Cloud marketplace, composed of different types of Clouds such as computing, storage, and content delivery Clouds, will be available to end-users and enterprises.

Users can interact with the Cloud market either transparently, by using applications that leverage the Cloud, or explicitly, by making resource requests according to application needs. At present, it is the responsibility of the users to directly interact with the Cloud provider. In the context of a real Cloud marketplace, users will indirectly interact with Cloud providers but they will rely on a market maker or meta-broker component, which is in charge of providing the best service according to the budget and the constraints of users. A Cloud broker client, directly embedded within applications, or available as a separate tool, will interact with the market maker by specifying the desired Quality of Service parameters through a Service Level Agreement. As a result of the query, the meta-broker will select the best option available among all the Cloud providers belonging to the Cloud marketplace.

Such interaction will take place through native interfaces exposed by the provider or via standardized brokering services.

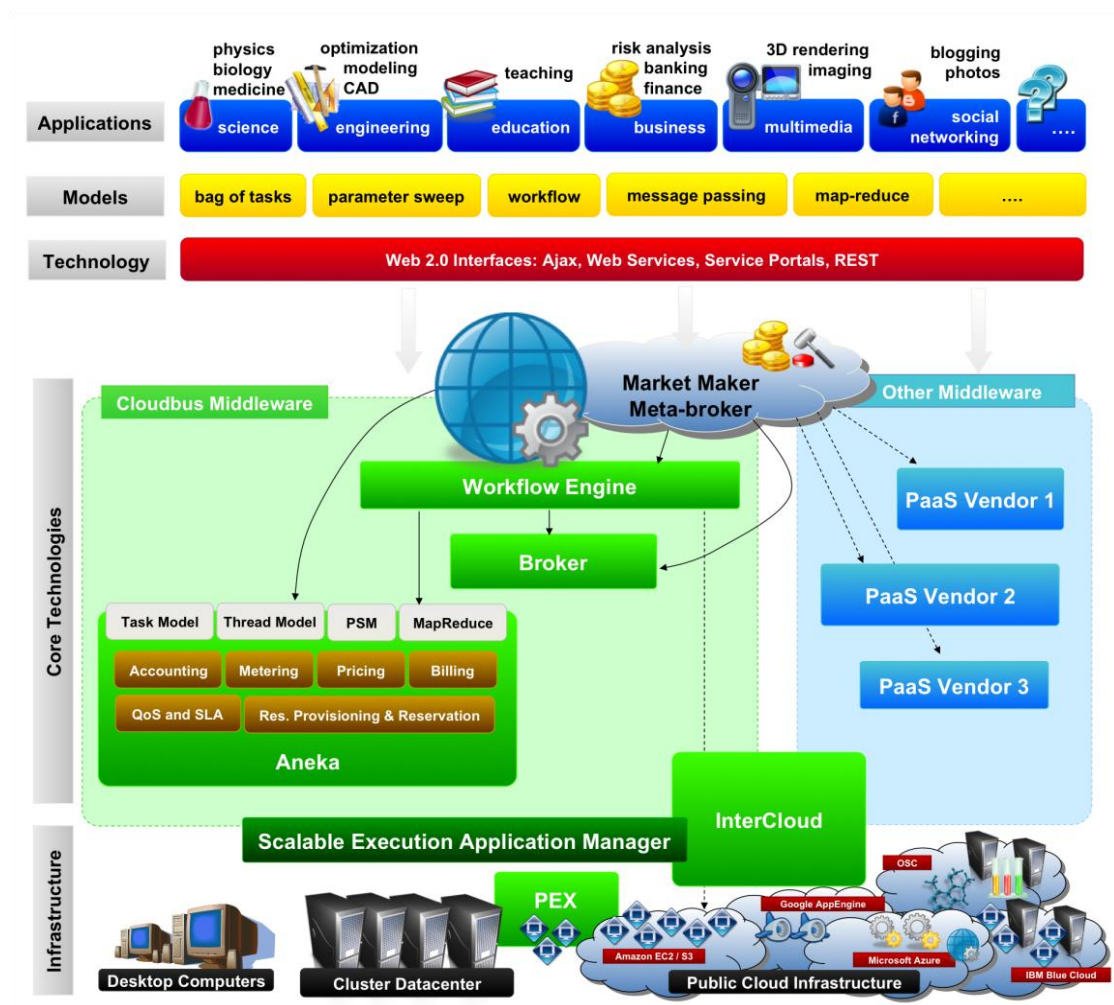

Figure 14.6: The Cloudbus Toolkit - A layered view of technologies and components for market oriented Cloud computing available within the Cloudbus toolkit.

In order to increase their chances of providing a better service to customers, different Cloud providers could establish peering arrangements among themselves in order to offload to (or serve from) other providers service requests. Such peering arrangements will define a Cloud federation and foster the introduction of standard interface and policies for the interconnection of heterogeneous Clouds. The integration of different technologies and solutions into a single value offering will be the key to the success of the Cloud marketplace. PaaS solutions, such as Aneka [43], could rely on different providers for leveraging the workload and balance the use of private resources by provisioning virtual resources from public Clouds. This approach applies for compute intensive, storage and content delivery. MetaCDN [6], which is a Content Delivery Cloud, aims to provide a unified access to different storage Clouds in order to deliver better service to end-users and maximize its utility.

The scenario projected by using the Cloud marketplace has its own challenges. Some of them have been already discussed in Section 14.2.4. In order to make this vision a reality, considerable amount of research has to be carried out through vigorous experiments. Simulation environments will definitely help researchers in conducting repeatable and controllable experiments, while devising new policies and algorithms for resource provisioning or new strategies for an effective and energy efficient use of physical resources. Simulation toolkit should support modeling of any possible scenario and any layer of the Cloud computing reference model: from the fundamental components of the infrastructure, such as physical nodes, data centers, and virtual machines, to the high level services offered to end users. This will help researchers to finely reproduce the problem frame they want to solve and to obtain reliable results.

The Cloudbus Toolkit is a collection of technologies and components that comprehensively try to address the challenges involved in making this vision a concrete reality. Figure 14.6 provides a layered view of the entire toolkit and puts it into the context of a real Cloud marketplace. At the top of the stack, real life applications belonging to different scenarios (finance, science, education, engineering, multimedia, and others) leverage the Cloud horsepower. Resources available in the Cloud are acquired by means of third party brokering services that mediate the access to the real infrastructure. The Cloudbus toolkit mostly operates at this level by providing a service brokering infrastructure and a core middleware for deploying applications in the Cloud. For what concerns the brokering service, the Market maker is the component that allows users to take full advantage of the Cloud marketplace. The Market maker relies on different middleware implementations to fulfill the requests of users: these can be Cloudbus technologies or third parties implementations. Figure 14.6 provides a breakdown of the components that constitute the Cloudbus middleware. Technologies such as Aneka and Workflow Engine provide services for executing applications in the Cloud. These can be public Clouds, private intranets, or data centers that can all be uniformly managed within an InterCloud [16] realm.

In the following sections, we will present more details about the Cloudbus toolkit initiative and describe how they can integrate with each other and existing technologies in order to realize the vision of a global Cloud computing marketplace.

## 14.4. CLOUDBUS / CLOUDS LAB TECHNOLOGIES

The CLOUDS lab has been designing and developing Cloud computing technologies to support science, engineering, business, creative media, and consumer applications. A summary of various Cloudbus technologies is listed in Table 14.2. In this section, we briefly describe each of these technologies.

Table 14.2: Components of Cloudbus Toolkit.

| Technology | Description |
|---|---|
| Aneka | A software platform for developing and deploying Cloud computing applications. |
| Broker | A middleware for scheduling distributed applications across Windows and Unix-variant distributed resources. |
| Workflow Management System | A middleware that handles dependent tasks, implements scheduling algorithms and manages the execution of applications on distributed resources. |
| Market Maker / Meta-Broker | A matchmaker that matches users' requirements with service provider's capabilities at a common marketplace. |
| InterCloud | A model that links various Cloud providers through peering arrangements to enable inter-Cloud resource sharing. |
| MetaCDN | A system that intelligently places users content onto "Storage Cloud" resources based on their QoS and budget preferences. |
| Data center Optimization | Adaptive allocations of compute, storage, and network resources to virtual machines and appliances. |
| Energy Efficient Computing | A research on developing techniques and technologies for addressing scalability and energy efficiency. |
| CloudSim | A simulation toolkit that helps users model: compute, storage, network and other related components of Cloud data centers. |

### 14.4.1 Aneka

Aneka [43] is a Platform-as-a-Service solution for Cloud computing and provides a software platform for developing and deploying applications in the Cloud. The core features of Aneka are: a) a configurable software container constituting the building blocks of the Cloud; b) an open ended set of programming models available to developers to express distributed applications; c) a collection of tools for rapidly prototyping and porting applications to the Cloud; d) a set of advanced services that put the horse power of Aneka in a market oriented perspective.

One of the elements that make Aneka unique is its flexible design and high level of customization allowing it to target different application scenarios: education, engineering, scientific computing, and financial applications. The Aneka container, which is the core of the component of any Aneka based Cloud, can be deployed into any computing resource connected to the Internet whether it be physical or virtual. This makes the integration with public and private Clouds transparent; and specific services for dynamic provisioning of resources are built into the framework in order to exploit the horse power of the Cloud. A collection of standardized interfaces, such as Web Services, make Aneka completely integrate with client applications and third party brokering services that can negotiate the desired Quality of Service and submit applications to Aneka Clouds.

### 14.4.2 Broker – Harnessing Cloud and other Distributed Resources

The Gridbus Resource Broker [44] is a market-oriented meta-scheduler for Computational and Data Grids, with support for a wide range of remote resource access services offered via various traditional middleware technologies such as Aneka [43], PBS, Globus, and SGE. It has been extended to provision compute and storage services offered by public Clouds such as Amazon EC2. The broker supports various application models such as parameter sweep, workflow, parallel tasks and bag of tasks.

The broker takes care of many functions that distributed applications require including discovering the right resources for a particular user application, scheduling jobs in order to meet deadlines and handling faults that may occur during execution. In particular, the broker provides capabilities such as resource selection, job scheduling, job management and data access to any application that requires distributed Grid resources for execution. The broker handles communication with the resources running different middleware, job failures, varying resource availability, and different user objectives such as meeting a deadline for execution or limiting execution within a certain budget.

The broker also provides QoS parameters in its service description for applications requiring a mix of public and private Cloud resources. Users specify QoS values for their applications at the brokers interface. The broker schedules the applications onto distributed resources comprising of local resources and Cloud resources to meet users QoS requirements. It facilitates dynamic provisioning policies where part of application workload can be moved to public Clouds and the remaining can be executed in the local resources. The division of workload is however dependent on the budget/deadline of the application and the capabilities of local resources to execute the application.

### 14.4.3 Workflow Engine

The Workflow Management System (WMS) [34] aids users by enabling their applications to be represented as a workflow and then execute on the Cloud from a higher level of abstraction. The WMS provides an easy-to-use workflow editor for application composition, an XML-based workflow language for structured representation, and a user-friendly portal with discovery, monitoring, and scheduling components. It can leverage Aneka [43] as well as the Gridbus Broker [44] to manage applications running on distributed resources. These tools put together enables users to select distributed resources on Clouds, upload/download huge amount of data to/from selected resources, execute applications on distributed resources using various scheduling algorithms and monitor the progress of applications in real-time.

A typical scenario is when the workflow engine is hosted as a PaaS and is using various other Cloud services, such as storage, content distribution, and so forth. In this case, users may submit applications via a web application running on Google App Engine, which delegates the requests to the workflow engine running in Amazon EC2. The application is billed using a SaaS application hosted by salesforce.com. Other enterprise network could use the workflow engine and submit jobs to Aneka

Enterprise Cloud.

Our recent focus has been on managing data intensive applications using the workflow engine in Clouds. We have developed several heuristics for scheduling workflow applications by leveraging distributed data retrievals. In contrast to using a single data source, we transfer segments of data from all available sources (to a compute resource for processing) in proportion to the cost of data transfer from the data storage locations. Hence, we schedule workflow tasks to resources and transfer data in order to minimize the total data transfer cost. We have a Particle Swarm Optimization (PSO-based), Non-linear Programming (NLP-based), and probe based heuristics for data retrievals and task scheduling.

Reliability of service providers is also one of the parameters that needs attention when using Cloud services. We have developed scheduling heuristics for workflow applications based on reliability. The WMS has been used for several real-world applications such as: fMRI brain imaging analysis [34], evolutionary multi-objective optimizations using distributed resources and intrusion detection systems with various models.

### 14.4.4 Market Maker/Meta-broker

Market Maker/Meta-broker [18] is a part of Cloud infrastructure that works on behalf of both Cloud users and Cloud service providers. It mediates access to distributed resources by discovering suitable Cloud providers for a given user application and attempts to optimally map the requirements of users to published services. The Market Maker is part of a global marketplace where service providers and consumers join to find suitable match for each other. It provides various services to its customers such as resource discovery, meta-scheduler, reservation service, queuing service, accounting and pricing services.

User application brokers send request for resources using the Cloud Exchange User Interface. The meta-broker discovers available resources and starts matching users' requirements (application broker for the meta-broker) to resource providers' capabilities. Upon suitable matching and reservation of resources, the user is notified for the available time slots. Users can directly use these resources for executing their jobs.

### 14.4.5 InterCloud

In the coming years, users will be able to see a plethora of Cloud several providers around the world desperate to provide resources such as computers, data, and instruments to scale science, engineering, and business applications. In the long run, these Clouds may require sharing its load with other Cloud service providers as users may select various Cloud services to work on their applications, collectively. Therefore, dispersed Cloud initiatives may lead to the creation of disparate Clouds with little or no interaction between them. The InterCloud model [16] will: (a) promote interlinking of islands of Clouds through peering arrangements to enable inter-Cloud resource sharing; (b) provide a scalable structure for Clouds that allow them to interconnect with one another and grow in a sustainable way; (c) create a global Cyber infrastructure to support e-Science and e-Business applications.

### 14.4.6 MetaCDN

MetaCDN [6] is a system that exploits "Storage Cloud" resources offered by multiple IaaS vendors, thus creating an integrated overlay network that provides a low cost, high performance CDN for content creators. It removes the complexity of dealing with multiple storage providers, by intelligently matching and placing the content provided by users onto one or many storage providers based on their quality of service, coverage and budget preferences. By using a single unified namespace, it helps users to harness the performance and coverage of numerous "Storage Clouds".

In the MetaCDN service, users can use the web portal or the SOAP and RESTful Web Services to: deploy content to geographically distributed locations as per their requirements; manage replica distribution according to their storage and data communication budget; view and modify existing distributed content.

### 14.4.7 Data center Optimization

Data centers form the core part of any Cloud infrastructure. They host compute, storage, network hardware where virtual machines are instantiated and leased to the users on demand. When allocating hardware to real-time/reserved requests, it becomes absolutely critical to use an adaptive algorithm such that the total cost of allocation is minimized. For instance, instantiating VMs in random machines may lead to too many machines being turned ON with least resource utilization. Likewise, if VMs are targeted to only few racks, "hot spot" could lead to higher power consumption, hardware overload etc. Thus allocation should take into account the QoS requirements of users and not over-provision resource on any single hardware. Allocation of virtual machines onto physical machines with multi-objective optimization is a challenging problem [45].

### 14.4.8 Energy Efficient Computing

In order to support elastic applications, Cloud infrastructure providers are establishing Data Centers in multiple geographic locations. These Data Centers are expensive to operate since they consume significant amount of electric power. For instance, the energy consumption of Google Data Center is equivalent to the power consumption of cities such as San Francisco. This is not only increasing the power bills, but also contributing to global warming due to its high carbon footprint. Indeed, the ICT sector is currently responsible for about 2 percent of global greenhouse gas emissions.

In our current research, we are investigating and developing novel techniques and technologies for addressing challenges of application scalability and energy efficiency with the aim of making a significant impact on industry producing service-oriented Green ICT technologies. As part of this, we explored power-aware scheduling [25], which is one of the ways to reduce energy consumption when using large data centers. Our scheduling algorithms select appropriate supply voltages of processing elements to minimize energy consumption. As energy consumption is optimized, operational cost decreases and the reliability of the system increases.

In a typical scenario, users send requests for VM provisioning to the global managers. The global managers exchange information for energy efficient VM allocation and migration. This information is shared with the local managers, which in turn control the VMs in the physical nodes.

### 14.4.9 CloudSim

The CloudSim toolkit [9] enables users to model and simulate extensible Clouds as well as execute applications on top of Clouds. As a completely customizable tool, it allows extension and definition of policies in all the components of the software stack. This makes it suitable as a research tool as it can relieve users from handling the complexities arising from provisioning, deploying, configuring real resources in physical environments.

CloudSim offers the following novel features: (i) support for modeling and simulation of large scale Cloud computing infrastructure, including data centers on a single physical computing node; and (ii) a self-contained platform for modeling data centers, service brokers, scheduling, and allocations policies. For enabling the simulation of data centers, CloudSim provides: (i) virtualization engine, which helps the creation and management of multiple, independent, and co-hosted virtualized services on a data center node; and (ii) flexibility to switch between space-shared and time-shared allocation of processing cores to virtualized services. These features of CloudSim would speed up the development of new resource allocation policies and scheduling algorithms for Cloud computing.

### 14.5. EXPERIMENTAL RESULTS

### 14.5.1 Aneka Experiment: Application Deadline driven Provisioning of Cloud Resources

The Cloud provides a market where compute resources can be leased by paying the usage cost. We used this concept and combined local and Cloud resources when executing an application with varying deadlines. Cloud resources were provisioned only when the local resources could not meet the application deadline. As a test application we used a bag of task application submitting 200 tasks with a variable deadline, each task taking 5 seconds to complete. We submitted this application with varying deadlines, as depicted in Table 14.3.

**Table 14.3: Using Cloud resources when required.**

| Deadline (Seconds) | Execution Time | Deadline Met? | Cloud Nodes Provisioned | Tasks on Local Resources | Tasks on Cloud (EC2) | Budget Spent (US$) |
|---|---|---|---|---|---|---|
| 60 | 144 | No | 18 | 115 | 85 | 0.765 |
| 120 | 118 | Yes | 9 | 112 | 88 | 1.530 |
| 180 | 170 | Yes | 5 | 125 | 75 | 0.425 |
| 1200 | 225 | Yes | 1 | 181 | 19 | 0.085 |

Table 14.3 clearly shows the advantage of using Cloud resources to meet the deadline when applications require tighter deadlines that cannot be fulfilled by local resources. Our scheduling algorithm estimated the total time the application would take when using local resources and only provisioned extra Cloud resources when this deadline could not be met. As we relaxed the deadline, from 60 seconds to 1200 seconds, the number of additional resources provisioned decreased from 18 to 1, and the cost of usage of EC2 resources from $US 0.765 to $US 0.085, respectively. Using Cloud as a readily available market, where compute resources are traded, we could meet the deadline for our test application.

### 14.5.2 Broker Experiment: Scheduling on Cloud and other Distributed Resources

The Gridbus broker's architecture and operational model is shown in Figure 14.7. As noted earlier, it supports market-oriented leasing of distributed resources depending on application and user's QoS requirements. At present, the broker can accommodate compute, storage, network, and information resources with prices based on compute time and bandwidth usage. It can also accommodate user objectives such as the fastest computation within the budget (time optimization), or the cheapest computation within the deadline (cost optimization) for both compute and data-intensive applications. The compute-intensive algorithms are based on those developed previously in Nimrod/G [7].

We have created a synthetic parameter sweep application (PSA) that executes a CPU intensive program with 100 different parameter scenarios or values. It led to the creation of an application containing 100 jobs, each job is modeled to execute ~5 minute with variation of (+/-20 sec.).

We have set the deadline of 40 minutes and budget of $6 for completing execution of the application; and conducted DBC (Deadline and Budget-Constrained) experiments for two different

optimization strategies:

- Time Optimization - this strategy produces results as early as possible, but before a deadline and within a budget limit.

- Cost Optimization - this strategy produces results by deadline, but reduces cost within a budget limit.

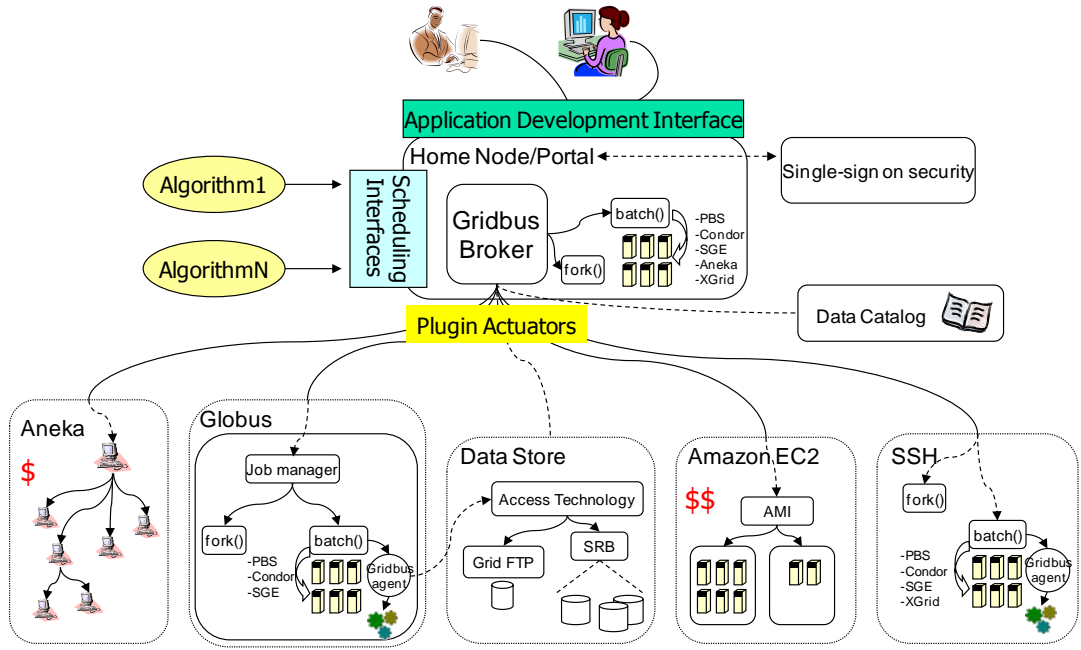

**Figure 14.7: A Resource Broker Architecture and Operational Model.**

We have used Grid and Cloud resources from Australia, Europe, and USA in these scheduling experiments. Table 14.4 shows resources details such as architecture, location, access price, and the number of jobs processed by them. These are shared resources and hence application scheduling has to be adaptive in nature. In the case of Amazon resources, the Broker on assigning jobs to them, created appropriate VM instances on EC2 resources and managed the execution of jobs on them. The access price (as Rate, indicated in cents/second; for readability purpose, it is multiplied by 1000) is based on a commodity market model. The access price used for EC2 resources is the actual price Amazon charges; whereas it is artificial for other resources. However, they assigned to reflect the offering of differentiated services at different costs as in the real-world marketplace.

Table 14.4: Grid and Cloud resources used in brokering experiments.

| Organization | Resource Details | Rate (Cents per second*1000) | Total Jobs Time-Opt | Total Jobs Cost-Opt |
|---|---|---|---|---|
| Georgia State University, US | snowball.cs.gsu.edu<br>8 Intel 1.90GHz CPU, 3.2 GB RAM, 152 GB HD, Linux | 90 (0.09) | 32 | 11 |
| H. Furtwangen University, Germany | unimelb.informatik.hs-furtwangen.de<br>1 Athlon XP 1700+ CPU, 767 MB RAM, 147 GB HD | 3 (0.003) | 4 | 5 |
| University of California-Irvine, US | harbinger.calit2.uci.edu<br>2 Intel P III 930 MHz CPU, 503 MB RAM, 32 GB HD | 2 | 8 | 10 |
| University of Melbourne, Australia | billabong.csse.unimelb.edu.au<br>2 Intel(R) 2.40GHz CPU, 1 GB RAM, 35 GB HD | 6 | 8 | 10 |
| University of Melbourne, Australia | gieseking.csse.unimelb.edu.au<br>2 Intel(R) 2.40GHz CPU, 1 GB RAM, 71 GB HD | 6 | 8 | 10 |
| Amazon EC2* | ec2-Medium instance<br>5 EC2 Compute Units*, 1.7 GB RAM, 350 GB HD | 60 | 14 | 16 |
| Amazon EC2* | ec2-Medium instance<br>5 EC2 Compute Units, 1.7 GB RAM, 350 GB HD | 60 | 13 | 16 |
| Amazon EC2* | ec2-Small instance<br>1 EC2 Compute Unit, 1.7 GB RAM, 160 GB HD | 30 | 7 | 11 |
| Amazon EC2* | ec2-Small instance<br>1 EC2 Compute Unit, 1.7 GB RAM, 160 GB HD | 30 | 6 | 11 |
| | Total Price / Budget Consumed | | 5.04$ | 3.71$ |
| | Time to Complete Execution | | 28 min | 35 min |

The results of scheduling experiments carried out using the Gridbus broker in May 2009 are summarized as follows:

| | Time Optimisation | Cost Time Optimisation |
|---|---|---|
| Budget Consumed | 5.04$ | 3.71$ |
| Time to Complete | 28 min | 35 min |

In the case of Time optimisation, the broker has scheduled jobs across all the available resources based on their completion rate even if they are costly as the as the job execution price is within the limit of budget available for each job. Many jobs are sent to powerful resources (such as one located in Georgia State University) from even if they are expensive as long as they are affordable and able to complete jobs quickly. For example, Georgia cluster is leased to process 32 jobs as it the most powerful resource. As a result, the broker is able to complete the application by 28 minute (much earlier than deadline) and spent $5.04.

In the case of Cost optimisation, the broker has preferred resources that are cheaper such as Amazon EC2 as long as they can complete assigned jobs within the deadline. It also scheduled jobs to other bit more expensive resources, just to make sure that deadline can be met. For example, expensive Georgia cluster is used to process only 11 jobs to ensure application execution by the deadline. As a result, the broker is able to complete the application by 35 minute (very close to the deadline) and spent $3.71, which is quite less compared to the amount spent by the Time optimization strategy.

These two experiments demonstrate that Cloud and Grid consumers can choose appropriate strategy for execution of their applications depending on timeframe by which results are needed. If the task at hand is priority/urgent one, they can choose the Time instead of the Cost optimization strategy. In addition, they can control the amount they are willing to invest for processing applications on market-oriented grid and cloud computing resources.

### 14.5.3 Deploying EGC Analysis Application in Cloud using Aneka

Advances in sensor technology, personal mobile devices, wireless broadband communications and Cloud computing are enabling real-time collection and dissemination of personal health data to patients and health-care professionals anytime and from anywhere. Personal mobile devices, such as PDAs and mobile phones, are becoming more powerful in terms of processing capabilities and information management and play a major role in people's daily lives.

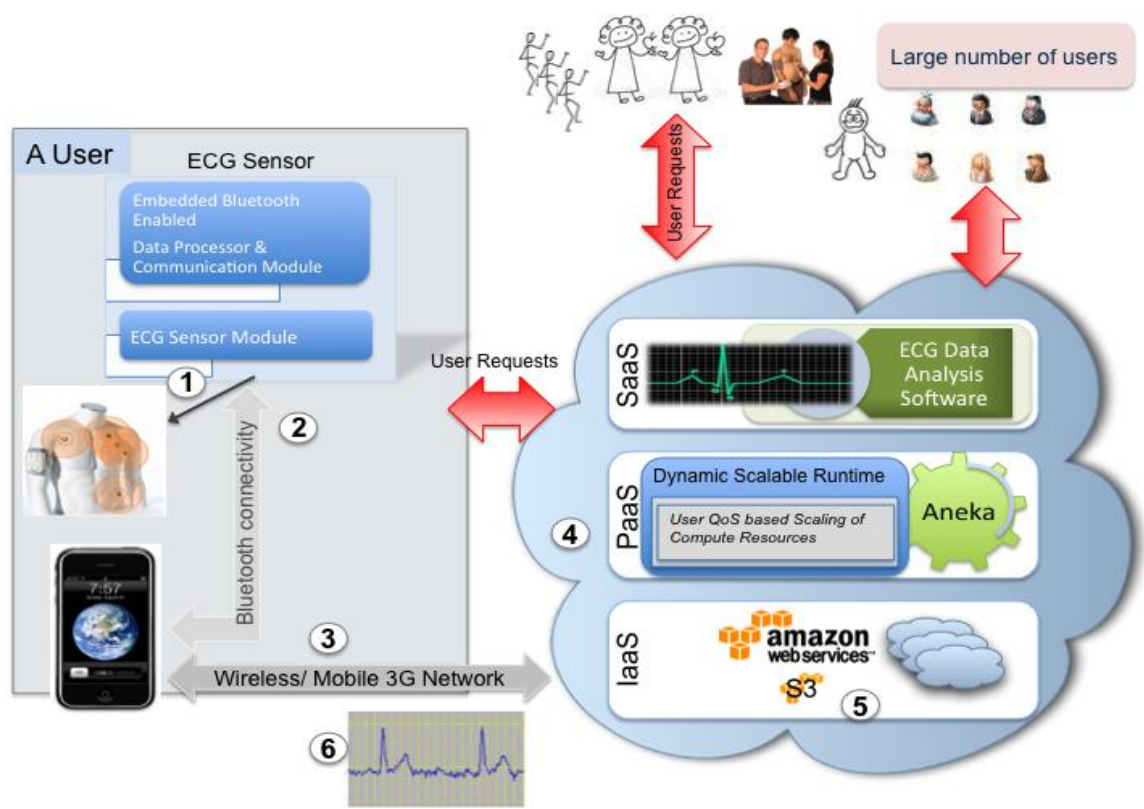

**Figure 14.8: Scaling Applications using Aneka Cloud computing technology.**

We designed a real-time health monitoring & computing system for people who suffer from cardiac arrhythmias. We implemented a personal health monitoring solution using real-time electrocardiogram (ECG) data to perform ECG beat and episode detection and classification. We developed a prototype system that collects people's electrocardiogram (ECG) data, disseminates them to information repository and facilitates analysis on these data using software services hosted in Clouds. We collected ECG data via sensors attached to a person's body, used a mobile device to communicate this data to the ECG analyzer (hosted as Software-as-a-Service in Clouds), disseminated the analyzed data to the person's and mobile phones when requested.

Figure 14.8 shows the components of the system. The environment hosting the health monitoring application must be able to handle large number of users, maintain user information and disseminate them as and when requested by them, as users pay for the services. The users possess a mobile device that is connected to a sensor device, which monitors the user's heartbeat. This mobile device communicates with the ECG data analysis service hosted in the Cloud to upload data and download the results in user readable format (e.g. graphs, statistical data, alerts, etc). The computation is carried out in the Cloud using services that can scale out depending on the number of user requests.

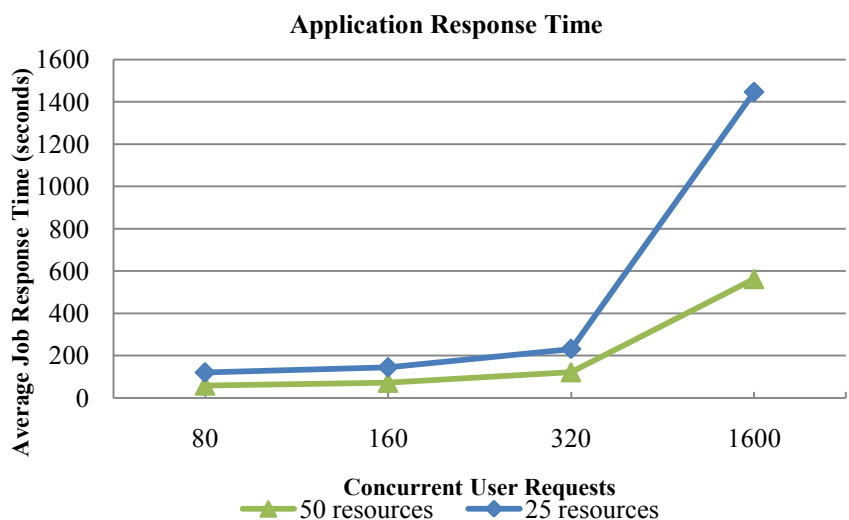

**Figure 14.9: Response time of ECG application for varying number of EC2 compute resources.**

The time taken by a system to respond to each user is of paramount importance in applications such as ECG monitoring. As the number of requests grows, the system suffers from increasing response time, as we show in Figure 14.9. When the maximum number of resources that can be used is limited to 25 resources, the response time for the ECG application depicted in Figure 14.8, increases as compared to the scenario when a maximum of 50 resources were used. This was mainly due to the queuing of user requests waiting for resources to be free. Even though the Cloud resources were instantiated dynamically, limiting the number of total resources used has significant effect on mission critical applications. Thus Cloud computing provides a platform for dynamically provisioning as many resources as and when required.

### 14.6. RELATED TECHNOLOGIES, INTEGRATION, AND DEPLOYMENT

The Cloudbus toolkit provides a set of technologies completely integrated with each other. More importantly, they also support the integration with third party technologies and solutions. Integration is a fundamental element in the Cloud computing model, where enterprises and end-users offload their computation to third party infrastructures and access their data anytime from anywhere in a ubiquitous manner.

Many vendors provide different solutions for deploying public, private, and hybrid Clouds. At the lowest level of the Cloud computing reference model, virtual server containers provide a management layer for the commodity hardware infrastructure: VMware , Xen [4], and KVM (Kernel-based Virtual Machine) are some of the most popular hypervisors available today. On top of these, Infrastructure-as-a-Service solutions such as Amazon EC2, Eucalyptus [33], and OpenNebula provide a high level service to end-users. Advanced resource managers such as OpenPEX and Haizea complete the picture by providing an advance reservation based approach for provisioning virtual resources on such platforms. Technologies such as Aneka and the Workflow Engine can readily be integrated with these solutions in order to utilize their capabilities and scale on demand. This applies not only for compute type workloads, but also for storage Clouds and CDNs, as demonstrated by the MetaCDN project. At a higher level, the Market maker and the Grid Service Broker are able to provision compute resources with or without a SLA by relying on different middleware implementations and provide the best suitable service to end-users.

The Cloudbus toolkit is a work in progress, but several Cloudbus technologies have been already put into action in real scenarios. A private Aneka Cloud has been deployed at GoFront (GoFront Group, Zhuzhou Electric Locomotive Works, is Chinas premier and largest nationwide research and manufacturer of rail electric traction equipment), in order to increase the overall productivity of product design and the return of investment of existing resources. The Workflow Engine has been used to execute complex scientific applications such as functional Magnetic Resonance Imaging (fMRI) workflows on top of hybrid Clouds composed of Amazon EC2 and physical clusters from labs worldwide. Various external organizations, such as HP Labs are using CloudSim for industrial Cloud computing research.

Furthermore, Aneka has been extended to support dynamic pooling of resources from public Clouds. This capability of Aneka enables creation of hybrid Clouds by leasing additional resources from external/public Clouds such as Amazon EC2 whenever the demand on private Cloud exceeds its available capacity. In addition, Aneka supports federation of other private Clouds within an enterprise, which are managed through Aneka or other vendor technologies such as XenServer and VMWare.

Moreover, some of our Cloudbus technologies have been utilized by commercial enterprises and they are demonstrated at public international events such as the 4th IEEE International Conference on e-Science held in Indianapolis, USA; and the 2nd IEEE International Scalable Computing Challenge hosted at the 9th International Conference on Cluster Computing and Grid (CCGrid 2009) held in Shanghai, China. These demonstrations included fMRI brain imaging application workflows, and gene expression data classification on Clouds and distributed resources.

## 14.7. SUMMARY AND FUTURE DIRECTIONS

In this paper, we introduced the fundamental concepts of market-oriented Cloud computing. We studied the building blocks of Cloud computing systems (IaaS, PaaS, and SaaS) and presented a reference model. The model together with the state of the art technologies presented in this paper, contribute significantly towards the mainstream adoption of the Cloud computing technology. However, any technology brings with it new challenges and breakthroughs. We detailed the major challenges faced by the industry when adopting Cloud computing as a mainstream technology as part of the distributed computing paradigm. We presented a utility-oriented Cloud vision that is a generic model for realizing market-oriented Cloud computing vision. Cloudbus realized this by developing various tools and platforms that can be used individually or together as an integrated solution. We also demonstrated via experiments that our toolkit could provision applications based on deadline, optimize cost and time of applications, and manage real-world problems via an integrated solution.

Market oriented computing in industry is getting real as evidenced by the plethora of vendors that provide Cloud computing services. For example, EC2 started with flat pricing then moved to pricing based on service difference and recently introduced auction based models. In the next two decades, service-oriented distributed computing will emerge as a dominant factor in shaping the industry, changing the way business is conducted and how services are delivered and managed. This paradigm is expected to have a major impact on service economy, which contributes significantly towards GDP of many countries, including Australia. The service sector includes health services (e-health), financial services and government services. With the increased demand for delivering services to a larger number of users, providers are looking for novel ways of hosting their application services in Clouds at lower cost while meeting the users quality of service expectations. With increased dependencies on ICT technologies in their realization, major advances are required in Cloud computing to support elastic applications offering services to millions of users, simultaneously.

Software licensing will be a major hurdle for vendors of Cloud services when proprietary software technologies have to be made available to millions of users via public virtual appliances (e.g. customized images of OS and applications). Overwhelming use of such customized software would lead to seamless integration of enterprise Clouds with public Clouds for service scalability and greater outreach to customers. More and more enterprises would be interested in moving to Clouds for cooperative sharing. In such scenarios, security and privacy of corporate data could be of paramount concern to these huge conglomerates. One of the solutions would be to establish a globally accredited Cloud service regulatory body that would act under a common statute for certifying Cloud service providers; standardizing data formats, enforcing service level agreements, handling trust certificates and so forth.

On one hand, there are technological challenges; on the other, there are issues with balancing usage cost and services delivered. Cloud service providers are already tussling by advertising attractive pricing policies for luring users of all kinds to use their services (e.g. Amazon, SalesForce, Google, etc.). As the market condition is determined through cutthroat competition between many vendors, dynamic negotiations and SLA management will play a major role in determining the amount of revenue to be generated for service providers. Similarly, users will be able to choose better services that fit their requirements and budget. They will be evaluating services based on their level of QoS satisfaction, so that they get the right value for the price paid.

As the price for commodity hardware and network equipments for a data center is already getting cheaper, significant part of the total cost of operating Cloud services in industrial scale is determined by the amount of energy consumed by the data center. To conserve energy and save cooling costs, data centers could adopt energy efficient resource allocation policies. Moreover, they could use renewable sources of energy to power up their centers and leave the least carbon footprint, in the long run. A daunting task for any vendor is to keep its Cloud services alive and running for as long as it takes. As users gradually become dependent on Cloud services, a sudden disruption of any of the services will send a ripple effect around the world that could: destabilizing markets (e.g. financial institutions such as banks depending on Clouds), paralyzing IT services (e.g. gmail services) and so forth. For preventing these effects arising from vendor "lock-in", interoperability issues between Cloud service providers should be adequately addressed. Nevertheless, Cloud computing is the technology for realizing a long awaited dream of using distributed compute, storage resources and application software services as commodities (computing utilities).

As the hype of Cloud computing matures and the technology gets adopted in the mainstream industry, the challenges and misunderstandings gradually get mitigated. State of various capabilities of Cloud computing noted in Gartner hype cycle released in July 2011 is shown in Figure 14.10. Overall, Cloud computing is still at the peak of the hype. Cloud application development tools, Cloud service integration and others are climbing the hype to reach the highest level of expectation, in the last 2 to 5 years. This hype curve represents trend in the industry and Cloudbus as an academic R&D project continue to advance the field of Cloud computing much earlier than what is noted in the hype curve. Hence, we believe that Cloudbus technologies are at the forefront of innovation in Cloud computing and its results will aid the industry in rapid progression of Cloud computing paradigm from technology trigger to plateau of productivity.

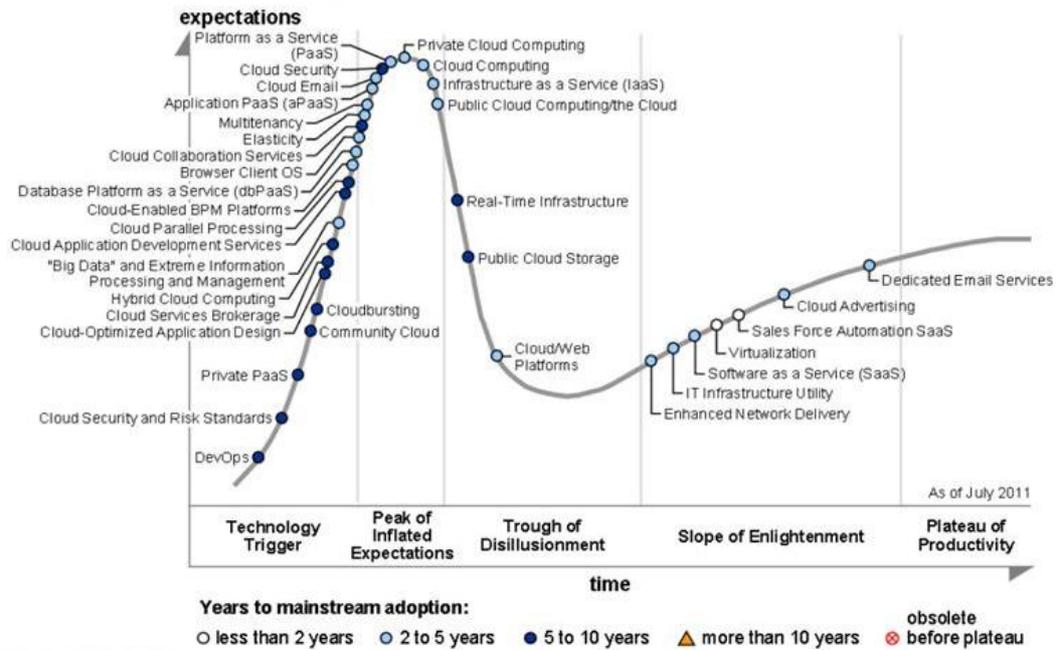

**Figure 14.10: Gartner 2011 Hype cycle of Cloud Computing.**

**ACKNOWLEDGEMENTS**

This work is partially supported through Australian Research Council (ARC) Discovery Project grant, and International Science Linkages (ISL) program of the Department of Innovation, Industry, Science and Research. All members of our CLOUDS Lab have been actively contributing towards various developments reported in this paper. In particular, we would like to thank Srikumar Venugopal, Xingchen Chu, Rajiv Ranjan, Chao Jin, Michael Mattess, William Voorsluys, Dileban Karunamoorthy, Saurabh Garg, Marcos Dias de Assunção, Alexandre di Costanzo, Mohsen Amini, James Broberg, Mukaddim Pathan, Chee Shin Yeo, Anton Beloglazov, Rodrigo Neves Calheiros, and Marco Netto. We also thank Mohsen Amini for providing us experimental results of execution of Broker on Clouds. This paper is a substantially extended version of the CloudCom 2009 conference paper [8].